# Discovering the influence of personal features in psychological processes using Artificial Intelligence techniques: the case of COVID-19's lockdown in Spain.

Blanca Mellor-Marsá, Alfredo Guitián[1], Andrew Coney[1], Berta Padilla and Alberto Nogales[2*]

[1] CEIEC Research Institute, Universidad Francisco de Vitoria, Ctra. M-515 Pozuelo-Majadahonda Km. 1,800, 28223 Pozuelo de Alarcón, Spain.
[*] Corresponding author: `alberto.nogales@ceiec.es`

**Abstract.**

Background / Introduction: At the end of 2019, an outbreak of a novel coronavirus (SARS-CoV-2) was reported in Wuhan, China, leading to the COVID-19 pandemic. In Spain, the first cases were detected in late January 2020, and by mid-March, infections had surpassed 5,000. On March 14, 2020, the Spanish government started a nationwide lockdown to contain the spread of the virus. While isolation measures were necessary, they posed significant psychological and socioeconomic challenges, particularly for vulnerable populations. Understanding the psychological impact of lockdown and the factors influencing mental health is crucial for informing future public health policies.

Methods: This study analyzes the influence of personal, socioeconomic, general health and living condition factors on psychological states during lockdown using Artificial Intelligence techniques. A dataset collected through an online questionnaire was processed using two workflows, each structured into three stages. First, individuals were categorized based on psychological assessments, either directly or in combination with unsupervised learning techniques. Second, various Machine Learning classifiers were trained to distinguish between the identified groups. Finally, feature importance analysis was conducted to identify the most influential variables related to different psychological conditions.

Results: The evaluated models demonstrated strong performance, with accuracy exceeding 80% and often surpassing 90%, particularly for Random Forest, Decision Trees, and Support Vector Machines. Sensitivity and specificity analyses revealed that models performed well across different psychological conditions, with the health impacts subset showing the highest reliability. For diagnosing vulnerability, models achieved over 90% accuracy, except for less vulnerable individuals using living environment and economic status features, where performance was slightly lower.

Conclusions: Depression is mainly influenced by health factors, while agoraphobia is shaped by living conditions. Economic factors play a key role in diagnosing vulnerability, but financial stability does not necessarily reduce psychological conditions.

# 1 Introduction

2020 will be remembered as the year of the coronavirus pandemic, caused by Severe Acute Respiratory Syndrome Coronavirus 2 (SARS-CoV-2). At the end of 2019, cases of unusual pneumonia were reported in Wuhan, China. On January 7, 2020, the virus responsible for the infection was identified as a novel coronavirus (Wang et al., 2020). This virus belongs to the same family as SARS-CoV, which caused a pandemic between 2002 and 2004, affecting over 8,000 individuals and resulting in 916 deaths, (Cherry & Krogstad, 2004). In this case, by March 2020, SARS-CoV-2 had already led to over 307,000 cases and more than 13,000 deaths, showing its significantly higher transmissibility compared to its predecessor.

The symptoms of COVID-19, as reported by (Grant et al., 2020), typically included fever, fatigue, cough, and dyspnea. SARS-CoV-2 exhibited a high level of transmission through human-to-human contact (Ahmed et al., 2018). In the absence of effective treatments or vaccines, self-isolation became the most effective measure to curb its spread (Livingston et al., 2020). The rapid transmission of the virus led governments and health authorities to implement varying degrees of lockdown measures (Shah et al., 2020). In Spain, a strict lockdown was imposed on March 14, 2020, experiencing different stages depending on the region lasting 99 days, (Mora et al., 2023). During this period, the population was required to stay home except for essential activities, such as work that could not be conducted remotely, accessing healthcare services or accompanying dependents.

These abrupt changes in social behaviors resulted in significant negative effects on individuals' mental well-being. Challenges such as separation from loved ones, disruptions to daily routines, fear of infection, and loss of personal freedom were commonly reported (Brooks et al., 2020). observed that enforced isolation significantly disrupted various aspects of life, often causing psychological stress and, in some cases, mental health issues. In Spain, (González-Sanguino et al., 2021) reveals significant increases in depressive symptoms during confinement. Another study, (Pedruzo et al., 2023), found that individuals with personality disorders had the highest anxiety and depression levels, but strict confinement did not significantly affect this group. Also (Pizarro-Ruiz & Ordóñez-Camblor, 2021) found that confinement caused emotional and behavioural changes in children and adolescents.

Mental health is influenced by a complex interplay of personal characteristics and broader social, economic, cultural, educational, environmental, and political factors, (Kirkbride et al., 2024). Among these, socioeconomic determinants, which include the conditions in which individuals are born, live, work, and age, play a critical role but are often overlooked in the literature. Psychological assessments provide valuable insights into the impact of various factors on mental health. When these assessments are combined with large datasets, Artificial Intelligence (AI) offers a powerful solution for managing and analyzing the data.

AI, a branch of computer science, seeks to replicate human intelligence in machines (Hoseini et al., 2018). Depending on the type of data, AI techniques can be divided into supervised and unsupervised learning approaches. Supervised learning involves training models using labeled data, while unsupervised learning identifies hidden patterns within unlabeled datasets, organizing information without explicit guidance (Santhya et al.). Supervised learning is often used for classification tasks, whereas unsupervised learning is commonly applied to clustering problems.

The main motivation of this study is to evaluate the impact of four sets of features on the mental health of the Spanish population during the 2020 COVID-19 lockdown. The study combines data related to personal characteristics, living conditions, economic or employment status and general health condition, with information assessing mental health outcomes. This data is processed using two different Machine Learning (ML) workflows comprising three stages. First stage, labels are assigned to the data, either directly through psychological tests or by combining them with unsupervised techniques. Second, classifiers are developed based on the datasets labelled in the first stage. Finally, a method is used to identify the input characteristics that most influence the ML models' classifications. As far as we know, the novelty of this paper lies on analyzing a particular dataset that describes the mental health of Spanish population during COVID's lockdown and, also, the application of different ML techniques to find how different features influenced these psychological conditions. As contributions, we can highlight the two different ML workflows that were applied and the set of most influential features on experiencing the psychological conditions that were obtained.

This paper is structured as follows: section 2 reviews previous studies on the influence of different factors on mental health during COVID-19. Section 3 provides a description of the data and methods used in this study. An analysis of the results can be found in section 4. Finally, a discussion of their implications is provided by section 5. The paper concludes with a summary of findings and suggestions for future research in section 6.

## 2     State of the art

Due to the high impact of the pandemic in the world, different works have been published during the last few years. This section tries to review most of the works that cover how the lockdown has impacted the mental health of diverse groups of human beings in different countries. Following, these papers are grouped as mental health issues in general, similar studies using ML techniques and those works focused on the Spanish population.

In the first group, we review the following studies. (Luijten et al., 2020) conducted a study with Dutch children and adolescents, comparing their mental and social health before the lockdown (December 2017–2018) and during it (April–May 2020). The features examined included anxiety, depressive symptoms, anger, and sleep-related impairment, analyzed using statistical methods such as PROMIS T-scores, ANCOVA, and multivariable linear regression. (Poudel & Subedi, 2020) assessed the impact of the lockdown on mental health and socioeconomic factors among the Nepalese population.

(Morgül et al., 2020) investigated the psychological effects on children and their caregivers in the United Kingdom during the summer of 2020, using statistical analyses and t-tests to derive their results. (Schwinger et al., 2020) focused on a German population to measure the impact of the lockdown on anxiety and depressive symptoms between late March and early June 2020, employing t-tests, Cohen's d, and ANOVA methods. (Alfawaz et al., 2021) studied the psychological effects of the lockdown on employees and students at a Saudi State University from May 11 to June 6, 2020. Their analysis utilized Chi-square tests, independent t-tests, and multinomial regression analysis. Butterworth (2022) leveraged Australian longitudinal data to evaluate the mental health effects of COVID-19 lockdowns, reporting a small overall decline but significant impacts on women with dependent children, which exacerbated preexisting inequalities. (Vacchiano, 2023) highlighted the disproportionate mental health impact of the first COVID-19 lockdown in Switzerland on younger generations, particularly Gen Y and Gen Z, attributing these effects to disruptions in social relationships and lifestyle routines. The study underscored the role of social capital and network mechanisms in understanding the disparities exacerbated by the lockdown. (Montero-Marin et al., 2023) found that the pandemic worsened the mental health of UK secondary students, particularly females and those initially at low risk, with protective factors including strong family ties, positive school environments, and supportive friendships. Finally, (Ferwana & Varshney, 2024) demonstrated that COVID-19 lockdowns in the United States significantly increased the use of mental health resources, with women and young adults experiencing the most substantial impacts. The study noted that mental health outcomes were more influenced by policy interventions than by the pandemic itself.

This group of studies conducts similar analyses but employs machine learning (ML) techniques. (Fiorillo et al., 2020) examined depressive, anxiety, and stress symptoms in the Italian population between March and May 2020, utilizing multivariate linear regression models to explore these features. (D'Urso et al., 2022) applied Support Vector Machines (SVM) and Random Forest (RF) to identify predictors of psychiatric symptom severity during the Italian COVID-19 lockdown. The models achieved up to 92% accuracy in predicting depression, anxiety, and obsessive-compulsive disorder (OCD) symptoms using pre-pandemic demographic and clinical data. Similarly, (Anbarasi et al. 2022) investigated the relationship between sleep and anxiety disorders during the COVID-19 pandemic lockdown in a cohort of 740 participants from India. RF was used to evaluate the influence of various attributes on sleep disorders, while K-Means clustered participants based on anxiety levels and sleep quality. An additional noteworthy study by (Shvetcov et al. 2023) analyzed university students across three COVID-19 time points: during lockdown, immediately after, and three months later. The analysis revealed three distinct subgroups—two with stable mental well-being and one significantly affected by traumatic stressors—highlighting time-dependent changes and burnout. Lastly, (Ntakolia et al., 2022) developed an explainable ML pipeline to assess the impact of the COVID-19 lockdown on the mood of children and adolescents in Greece with pre-existing mental health conditions. Their approach combined clustering techniques such as K-Means and Clustering Linkage with classification models like Logistic Regression and SVM. SHAP (SHapley Additive exPlanations) was then used to determine the influence of specific features on model predictions. When comparing these studies with the present work, we observe that while they employ similar ML methodologies, the present study focuses on a distinct population and evaluates the impact of a different set of features.

Among the studies examining the impact of the COVID-19 lockdown in Spain, the following works stand out. (Tubío-Fungueiriño et al., 2022) used machine learning (ML) to predict changes in obsessive-compulsive disorder (OCD) symptoms among 127 patients during the pandemic. Although anxiety and depression symptoms were predicted with lower reliability, the findings suggest that this approach could assist clinicians in identifying high-risk patients and optimizing care strategies. (Aperribai et al., 2020) explored the role of physical activity in mitigating the mental health impact on teachers, focusing on work, family, and social relationships. This study gathered data from primary and secondary school teachers in Spain and applied descriptive statistics, Cronbach's alpha, and t-tests to analyze the results. (Jojoa et al., 2022) analyzed the effects of COVID-19 lockdowns on university students and staff in Spain and Colombia, employing natural language processing techniques for sentiment analysis. Their findings revealed that negative sentiments dominated in both countries, with Spanish respondents expressing more pronounced negative emotions, while also highlighting disparities in infrastructure and educational challenges during the pandemic. (Benito et al., 2024) applied machine learning models to predict anxiety, depression, and self-perceived stress among 9,291 individuals in Catalonia after the COVID-19 pandemic. By integrating explainable AI techniques such as SHAP and UMAP, they identified key risk factors—such as poor social support and chronic health conditions—and stratified populations into high-risk profiles to support targeted mental health interventions. (Ryu et al., 2021) investigated changes in social media usage patterns during the COVID-19 lockdown in Madrid, Spain, and their association with clinical anxiety symptoms. Their findings, based on machine learning models, revealed that patients with severe anxiety were less active on communication apps but more engaged with social network apps, suggesting that these distinct digital behaviors could serve as biomarkers for anxiety. Finally, (Solano et al., 2021) examined changes in smartphone usage patterns in Spain before and after the first COVID-19 lockdown in 2020, using machine learning techniques such as clustering and Isomap projections. The study identified four distinct user profiles, highlighting significant increases in communication and social app usage during the lockdown, particularly among previously low-usage groups, and explored correlations with sociodemographic, behavioral, and psychological factors Although different works are studying the impact of COVID lockdown on Spanish mental health none of them accomplishes the characteristics of the actual approach.

Summarizing, the present work provides some novelties compared with the previous work. First, the studied population corresponds to a homogeneous set of people living in Spain during the lockdown which was strict compared with other countries. Second, the main aim is not only to measure the impact on mental health but also to study how this is influenced by different personal, living, economic and general health features. Finally, this is the only work that uses ML techniques to perform the work for this population during unique circumstances.

# 3 Materials and methods

In this section, we describe the data used to experiment and how it has been collected. A theoretical formalisation of the methods used in the analysis is also given.

## 3.1 Data compilation

In this study, we utilized the publicly available dataset LOCKED[1], (Mellor-Marsá et al., 2024). Data for this research was collected online through a questionnaire designed with two main sections: personal data and a psychological test. The questionnaire (see Appendix A) was distributed using Google Forms and promoted via various social networks and research mailing lists. All participants provided informed consent for their data to be used in this study.

Due to the lockdown situation, participants completed the questionnaire independently, without supervision from experts. Inclusion criteria required participants to be at least 18 years old and residents of Spain. The questionnaire was launched on May 2, 2020, and data collection concluded on May 9, 2020. This marked the start of "stage 0" in some regions of Spain, during which residents were allowed to walk outdoors within specific time slots, being the end of strict isolation measures.

The final dataset comprised responses from 981 individuals with diverse profiles. It included 96 items, of which 51 were categorized as personal, environmental, economic, and general health features, while the remaining 45 assessed general psychopathology through the psychological test.

**Compilation of personal features.** The first part of the questionnaire is used as a set of independent features. This means that this set of features will be used to know which are the ones that influence the most in the studied psychological conditions.

Demographics: genre, age, civil status, place of birth, identity documents, nationality, education level and family structure.

Living Environment: type of housing, home characteristics, flatmates and number of social contacts before and after the quarantine.

Economic Status and Employment: work situation before and after the quarantine, type of work, hours of dedication before and after quarantine, incomes before and after quarantine, capacity to assume the monthly expenses and debts.

Health and COVID-Related Impacts: general health conditions and the specific impact of COVID-19 on health and caregiving.

**Psychological conditions test.** In the second part of the data collection, the Spanish version of the Symptom Assessment-45 Questionnaire (SA-45) was used, (Holgado Tello et al., 2019). The SA-45 is a self-administered questionnaire consisting of 45 items (Appendix B) that yield a general psychopathology score along with measures for nine symptom dimensions: obsession-compulsion, interpersonal sensitivity, hostility, anxiety, somatisation, paranoid ideation, phobic anxiety, psychoticism and depression, each grouped into sets of five items. Responses are scored on a Likert scale from 0 ("Not at all") to 4 ("Very much or extremely"), with total scores ranging from 0 to 180 points and dimension scores from 0 to 20 points; higher scores indicate greater levels of psychopathological symptomatology. The sum of the scores for the items within each scale provides a scale-specific score for each of the nine dimensions that is used to label each individual.

---

[1] https://zenodo.org/records/14203988

**Data cleaning.** During data cleaning, outliers, and individuals with insufficient or poor-quality information were removed, and missing values were imputed, leaving a dataset of 979 instances. Open-ended variables with high variability, like place of birth, postal code, or specific job, were excluded due to their broad scope and the noise they could introduce. The remaining variables were encoded for model estimation. The final set regarding the independent features' dataset comprises 41 features, grouped into the four subsets described above personal data, living conditions, economic features and general health condition. Appendix C shows the final distribution of the features depending on the four subsets.

### 3.2 How distinctive features influence psychological conditions.

This work aims to obtain what features influence the appearance of different psychological conditions in the population that experienced the strict lockdown that took place in Spain due to the COVID-19 spread. In this case, a three-step pipeline that benefits from the application of AI technique has been implemented as follows. First, there is a need to label the individuals based on a psychological test. This has been approached in two ways: the simpler one corresponds to the test itself and the other involves the use of unsupervised methods combined with the psychological test values. During the second stage, different supervised classification methods were applied to the different subsets obtained in the data-cleaning stage. As two different labelling methods were applied, classifiers can be applied to different use cases for the diagnosis of psychological conditions. Finally, once we have obtained accurate classifiers for the four subsets and the two labelling methods, we use interpretability techniques so we can obtain which of the compiled features are the ones that influence more in discriminating between classes. This workflow is described in Fig. 1.

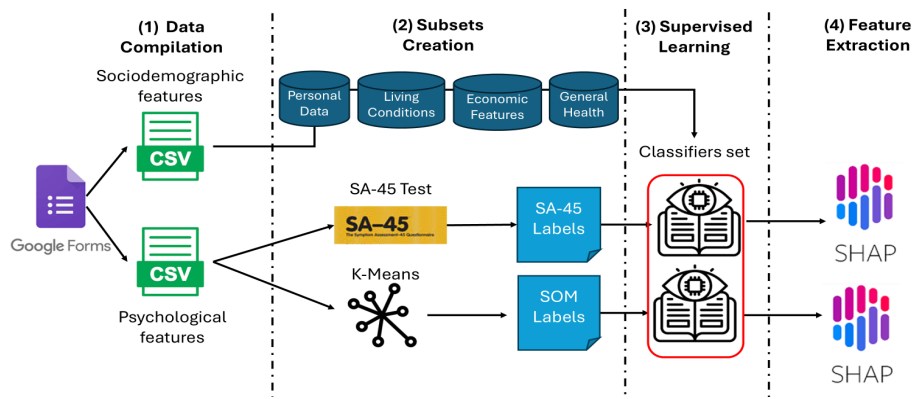

**Fig. 1.** Workflow to obtain the most influential features for psychological conditions.

As can be seen above, the workflow comprises a set of supervised and unsupervised methods and another for ML interpretability. Following, we formalise all the methods used in the three steps.

**Unsupervised models.** First, as classifiers used labelled data, we need to give a class to every individual of the dataset. These classes are based on the SA-45 test leading to two use cases. The first uses just the nine classes given by the test and the other uses an unsupervised learning clustering method called K-Means. In this method, we will cluster the dataset using the responses to the questions in the test and then will give a descriptive name to each of the clusters based on the nine psychological conditions from the SA-45 test.

*K-Means.* This clustering algorithm works as follows, (Jin & Han, 2011). Starting with an initial but suboptimal clustering, each point is reassigned to its nearest centre, and the clustering centres are updated by calculating the mean of the points assigned to each centre. This relocating and updating process is repeated until the convergence criteria, such as a predefined number of iterations or a change in the value of the distortion function, are met.

**Supervised models.** In the second stage of the workflow, we are using the two labelled datasets obtained in the previous stages to train different approaches of supervised classifiers. In total, we have used seven different models that are formalized following.

*Decision Trees.* This model employs a "divide and conquer" strategy for making classifications. The algorithm constructs a tree structure where each node corresponds to a feature of the dataset, each branch signifies a decision, and each leaf indicates an outcome or class, (Breiman et al., 2017).

*Random Forest (RF).* It is an ensemble learning method that constructs multiple decision trees during training and combines their class assignments to enhance accuracy and reduce overfitting, (Breiman, 2001). Each tree in the forest is trained on a different sample of the dataset (with replacement) and utilizes a random subset of features.

*Support Vector Machines (SVM).* (Cortes & Vapnik, 1995) designed this model based on identifying the hyperplane that most effectively separates different classes in a high-dimensional space. The algorithm works by maximizing the margin between the samples of the closest classes to the hyperplane.

*Gradient Boosting.* Introduced by (Friedman, 2002) is an ensemble method that constructs a strong predictive model by combining weaker models, usually decision trees. Each model is built sequentially, with adjustments made to its weights to correct the errors of the classification model, thereby minimizing the loss function.

*Naïve Bayes.* The Naïve Bayes classifier relies on the simplifying assumption that the attribute values are conditionally independent given the target value, (Mitchell, n.d.). In other words, it assumes that, given the target value of an instance, the probability of observing the conjunction of attributes is simply the product of the individual attribute probabilities.

*Logistic Regression (LR).* Logistic Regression is a linear classification model designed to predict the probability of a specific class using a logistic function, (Mccullagh, 2008). While it is linear in its fundamental form, it is easy to interpret and can be extended to accommodate nonlinear data by incorporating polynomial terms or interaction effects.

*Multilayer Perceptron (MLP).* Introduced in (Rumelhart et al., 1986), it is a type of artificial neural network that comprises multiple layers of nodes (neurons), including an input layer, one or more hidden layers, and an output layer. Each node is connected to nodes in the subsequent layer through weighted connections, which are adjusted during training to minimize prediction errors.

**Models' interpretability.** The last step is the one aimed at obtaining the features that influence acquiring different psychological conditions. A way to carry it out is that of Shapley Additive Explanations (SHAP) values which identify the most key features in each of the three labelling use cases applied to the three subsets: personal data, living conditions and economic features. By evaluating the contribution of each feature to the model's classifier, SHAP offers a consistent measure of feature importance, helping to interpret complex ML models, (Arthur & Date, 2021) & (Lundberg et al., n.d.). This method enhances understanding of the key factors influencing mental health during lockdown, providing valuable insights into the data.

**Evaluation metrics.** To provide a thorough evaluation and gain detailed insights into model performance, multiple metrics were employed, following are those related to the unsupervised models.
Error of inertia is the measure of the total within-cluster variance, representing the sum of squared distances between data points and their nearest cluster centroid. In the context of K-Means clustering, this is the value that the algorithm minimizes during optimization, (Arthur & Date, 2021). This metric is described in Equation 1.

$$Error_{Inertia} = \sum_{i=1}^{n} \min_{i \leq k \leq K} \|x_i - c_k\|^2 \tag{1}$$

In the context of K-Means the equation above is interpreted as following. n represents the number of data points in the dataset, and $K$ denotes the number of clusters. Each data point is represented as $x_i$, and $c_k$ refers to the centroid of the $k^{th}$ cluster. The squared Euclidean distance, $\|x_i - c_k\|^2$, measures the distance between a data point $x_i$ and its assigned cluster centroid $c_k$, and this distance is used to calculate the inertia, which represents the total within-cluster variance.

Regarding the metrics for the supervised methods, we are using the followings. Accuracy, which measures the proportion of correct predictions, served as the primary metric across the different classifiers. Equation 2 describes this metric.

$$Accuracy = \frac{TruePositives\ (TP) + TrueNegatives\ (TN)}{TruePositives\ (TP) + TrueNegatives\ (TN) + FalsePositives\ (FP) + FalseNegatives\ (FN)} \tag{2}$$

In addition to Accuracy, models were assessed using Sensitivity and Specificity. Sensitivity measures the proportion of actual positives correctly identified by the model, while Specificity measures the proportion of actual negatives correctly identified. These metrics are critical for understanding the types of errors the model makes, particularly when FP or FN carry excessive costs. Equation 3 is used to formalize Specificity and Equation 4 for Sensitivity.

$$Specificity = \frac{TN}{TN + FP} \tag{3}$$

$$Sensitivity = \frac{TP}{TP + FN} \tag{4}$$

## 4 Results and discussion

This section covers the obtained groups after labelling the individuals with the SA-45 and the unsupervised models. Also, it covers all the metrics used to obtain accurate classifiers. Finally, the most influential features are compiled in different tables.

### 4.1 Obtained datasets after the labelling process.

**SA-45 labelling.** As we have said before the first step of the proposed method consists of labelling individuals based on the psychological conditions provided by the SA-45 test. The test can evaluate nine different psychological conditions apart from not experiencing any and belonging to the control group. Table 1 compiles how individuals are distributed among classes, we should highlight that an individual can be in more than one class that is why the sum up of the cases among psychological conditions is greater than the total amount of individuals.

**Table 1.** Number of cases per psychological process group following the SA-45 test.

| Psychological process | Number of cases |
|---|---|
| Hostility | 146 |
| Somatization | 158 |
| Depression | 158 |
| Obsession-Compulsion | 167 |
| Anxiety | 169 |
| Interpersonal Sensitivity | 169 |
| Agoraphobia | 124 |
| Paranoid Ideation | 130 |
| Psychoticism | 140 |
| Control (no class) | 620 |
| **Total** | **981** |

Apart from labelling individuals directly with the test, we have applied an unsupervised learning method called K-Means.

**K-Means labelling.** In the case of K-Means, we applied the algorithm to a subset consisting solely of the features derived from the psychological test, SA-45. A key decision in this algorithm is determining the optimal number of clusters to form the new groups. This decision was informed by several analyses. Figure 2 that which is called "elbow graph" and shows from which cluster the error is stable, so we can use it as a first point to take the decision. This is complemented with Table 2 that presents the error of inertia for different numbers of clusters, helping to identify the optimal number by detecting the point of diminishing returns, often referred to as the "elbow point". Figure 2 provides a line diagram that illustrates how the nine psychological conditions measured by the SA-45 test behave across clusters, enabling a descriptive interpretation of each cluster.

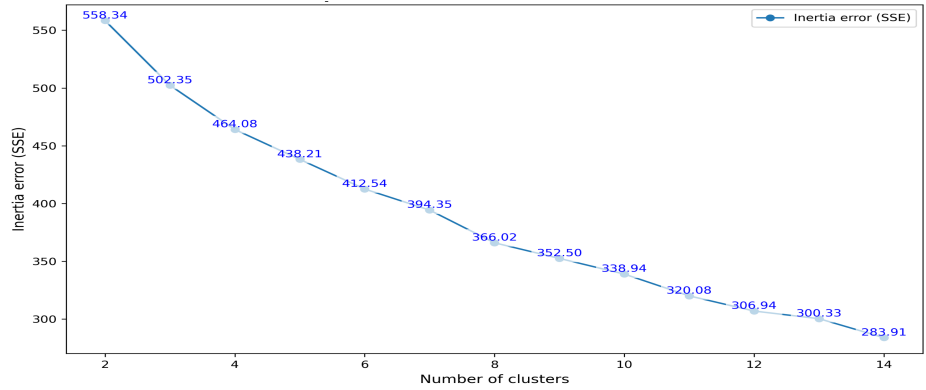

**Fig. 2.** Error inertia after applying K-Means

Regarding Figure 2 that shows the value of the inertia error for each number of clusters, we can delimit the lower k in 3 or 4 clusters. We complemented the information provided in the previous figure with data compiled in Table 2. In this table, column 3 represents the difference in inertia error between a cluster and the subsequent one. From this data, it is evident that the error begins to stabilize starting from cluster 4. However, cluster 3 should not be dismissed as the optimal choice, as it marks the point where the differences in error fall within a relatively small range, indicating a possible balance between compactness and simplicity in the clustering solution.

**Table 2.** Error inertia for K-Means

| Number of clusters | Error of inertia | Error difference |
|---|---|---|
| 2 | 558.34 | 0.00 |
| 3 | 502.35 | 55.99 |
| 4 | 464.08 | 38.27 |
| 5 | 438.21 | 25.87 |
| 6 | 412.54 | 25.67 |
| 7 | 394.35 | 18.19 |
| 8 | 366.02 | 28.33 |
| 9 | 352.50 | 13.52 |
| 10 | 338.94 | 13.56 |
| 11 | 320.08 | 18.87 |
| 12 | 306.94 | 13.13 |
| 13 | 300.33 | 6.61 |
| 14 | 283.91 | 16.42 |

Given that four clusters appear to be a suitable choice, we created a line diagram illustrating how frequently each psychological condition has been diagnosed within each cluster. To account for differences in the number of individuals in each cluster, the frequency of each condition was normalized by dividing it by the number of individuals in the respective cluster. Fig. 3 provides a visual representation of how the clusters differ in their behaviour with respect to the psychological test, offering insights into the distinctive patterns of each cluster.

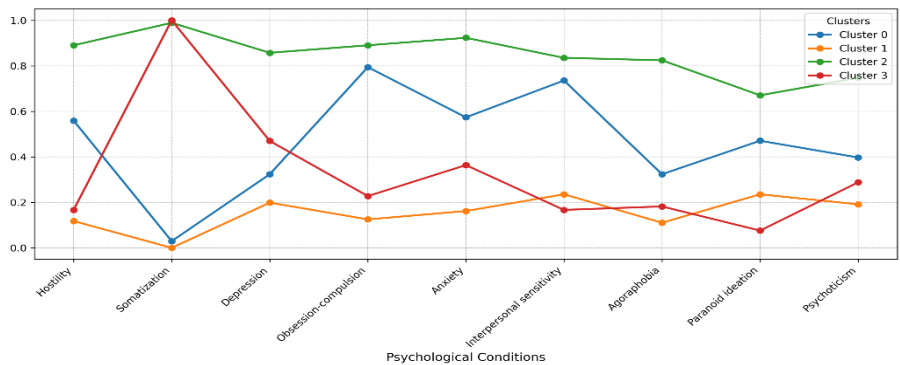

**Fig. 3.** Diagram of lines representing the clusters' behaviour for k=4.

As observed, the two upper clusters appear to exhibit different behaviour compared to the rest. Clusters 1 and 3, on the other hand, seem to behave very similarly. Considering this similarity, we considered merging them. To make this decision, we calculated the distances between the centroids of each cluster. If clusters 1 and 3 are determined to be sufficiently close based on these distances, we can merge them to create a single, new cluster. Table 3 shows this information, which can guide decisions about merging similar clusters to refine the overall clustering solution.

**Table 3.** Distances between clusters' centroids

| Cluster | Distance to centroid 1 | Distance to centroid 2 | Distance to centroid 3 | Distance to centroid 4 |
|---|---|---|---|---|
| 1 | 0.00 | 1.11 | 1.37 | 1.41 |
| 2 | 1.11 | 0.00 | 2.13 | 1.08 |
| 3 | 1.37 | 2.13 | 0.00 | 1.69 |
| 4 | 1.41 | 1.08 | 1.69 | 0.00 |

As observed above, cluster 1 is the closest one to cluster 3. Based on this, we decided to merge these clusters, reducing the total number of clusters to three. Following this adjustment, we created a new line diagram to evaluate whether the differences in behaviour among the clusters have become more pronounced, Fig. 4.

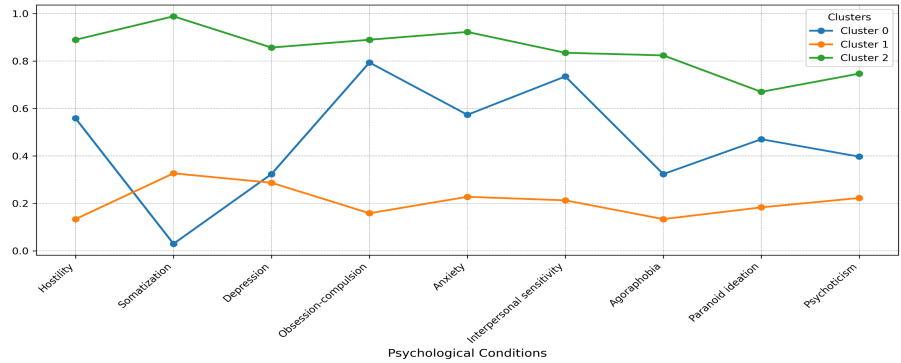

**Fig. 4.** Diagram of lines representing cluster's behaviour after merging clusters 1 and 3.

As shown in the figure above, using three clusters instead of four makes the behavior of each cluster more distinct. In this configuration, Cluster 2 clearly stands out in all the psychological conditions compared to the other clusters. Cluster 0 exhibits medium-range values for most features when compared to Cluster 1, except for the feature "Somatization," where it differs. Finally, Cluster 1 consistently shows the lowest values, except for "Somatization," and is generally characterized by very low values across all features.

Based on these observations, we can describe the clusters in terms of vulnerability levels. Cluster 2 represents individuals with high vulnerability, Cluster 0 corresponds to individuals with average vulnerability, and Cluster 1 consists of individuals with low vulnerability.

To complement this analysis, we account for the fact that individuals may experience more than one psychological condition. We classify individuals into similar categories based on their degree of vulnerability. Considering the nine psychological conditions, individuals experiencing one to three conditions are labeled as having low vulnerability, those with four to six conditions are categorized as vulnerable, and those experiencing seven or more conditions are deemed highly vulnerable. The percentage distribution of individuals across these categories is presented in Table 4.

**Table 4.** Distribution of vulnerability level among clusters

| Cluster | Group description | Number of individuals | Percentage |
|---|---|---|---|
| 0 | **Vulnerable** | **52** | **76.47 %** |
|   | Less vulnerable | 16 | 23.53 % |
| 1 | **Less vulnerable** | **178** | **88.12 %** |
|   | Vulnerable | 24 | 11.88 % |
| 2 | **Highly vulnerable** | **74** | **81.32 %** |
|   | Vulnerable | 17 | 18.68 % |

15The individuals in each group provide a clear basis for defining the characteristics of each cluster. Moreover, this classification aligns well with the previous interpretation of the line diagram in Figure 4. Based on this analysis, we assign the following labels to individuals in our subsets: less vulnerable, vulnerable, and highly vulnerable. Additionally, we include the label control for instances that have not experienced any psychological condition. Table 4 shows the final distribution of individuals for each class.

### 4.2 Trained classifiers

Once we have labelled the individuals with two strategies, two use cases are generated for the classification stage. For all approaches, we are splitting the sociodemographic features into four subsets. This entails building seven classifiers for seven different ML approaches applied to the different subsets.

As good practices for training the models, we have applied the following strategies. The datasets were divided into training, validation, and test sets. The training and validation sets were used to fine-tune model hyperparameters, while the test set evaluated the model's final performance. An 80-20% split was applied for the training, validation and test sets, randomizing instances to avoid biases. A grid search strategy that explores various combinations was used to optimize the hyperparameters during training, (Bergstra & Bengio, 2012). Table 5 compiles all the hyperparameters used in the different ML models during the application of this strategy.

**Table 5.** Hyperparameters used during grind search.

| ML model | Hyperparameters |
| --- | --- |
| Decision tree | Maximum depth: [None, 10, 20, 30] |
|  | Minimum samples split: [2, 5, 10] |
|  | Minimum samples per leaf: [1, 2, 4] |
|  | Split Criterion: ['Gini', 'Entropy'] |
| RF | Number of estimators: [50, 100, 150] |
|  | Maximum depth: [None, 10, 20] |
|  | Minimum samples split: [2, 5] |
|  | Minimum samples per leaf: [1, 2] |
| SVM | Regularization parameter (C): [0.1, 1, 10] |
|  | Kernel: ['Linear', 'Radial Basis Function (RBF)', 'Polynomial'] |
|  | Degree (for Polynomial kernel): [3, 4, 5] |
|  | Gamma: ['Scale', 'Auto'] |
| Gradient Boosting | Number of estimators: [50, 100, 150] |
|  | Learning rate: [0.01, 0.1, 0.2] |
|  | Maximum depth: [3, 4, 5] |
|  | Subsampling ratio: [0.8, 1.0] |
| Naïve Bayes | Variance smoothing: [1.0 … 1e-9] |
|  | Alpha: [0.1, 0.5, 1.0, 1.5, 2.0, 5.0] |
|  | Class prior estimation: [True, False] |
| LR | Regularization parameter (C): [0.01, 0.1, 1, 10, 100] |
| MLP | Regularization type: [L1, L2]<br>Optimization algorithm: ['Library for Large Linear Classification', 'Limited-memory Broyden–Fletcher–Goldfarb–Shanno (LBFGS)']<br>Maximum iterations: [100, 200, 500]<br>Hidden layer sizes: [(50,), (100,), (50, 50)]<br>Activation function: ['ReLU', 'tanh']<br>Solver: ['Adam', 'Stochastic Gradient Descent (SGD)']<br>Alpha: [0.0001, 0.001]<br>Learning rate type: ['Constant', 'Adaptive'] |

Finally, k-fold validation has been used as a method to validate the models more thoroughly. Since some ML models initialize their hyperparameters randomly, this initialization, combined with the data split, could positively influence the model's performance.

**SA-45 labels.** In the first use case, the psychological test has been directly applied generating ten labels which are the nine psychological conditions plus the control group. Starting with this data, we apply an approach of binary classification to know if an individual is experiencing one of the psychological conditions or not.

This leads to the results compiled in Table 6 where we find for each of the four subsets the accuracy in train, validation, and test for each of the nine classifiers. The values are represented with the mean average and standard deviation due to the use of the k-fold validation.

**Table 6.** Accuracy for the ML classifiers using SA-45 labels.

| Training subset | Binary classifier | Best ML model | Train | Validation | Test |
|---|---|---|---|---|---|
| Demographic characteristics | Hostility | DT | 91.15% ± 2.50 | 85.16% ± 1.48 | 81.36% |
| | Somatization | SVM | 81.70% ± 5.86 | 76.36% ± 3.09 | 87.50% |
| | Depression | SVM | 85.09% ± 4.47 | 80.27% ± 2.85 | 87.50% |
| | Obsession-Compulsion | RF | 90.96% ± 2.40 | 82.40% ± 1.50 | 89.55% |
| | Anxiety | SVM | 81.89% ± 5.42 | 77.81% ± 3.47 | 89.71% |
| | Interpersonal Sensitivity | DT | 89.60% ± 2.42 | 87.59% ± 1.01 | 85.29% |
| | Agoraphobia | RF | 92.76% ± 1.25 | 85.89% ± 0.53 | 90.00% |
| | Paranoid Ideation | DT | 92.23% ± 1.69 | 84.52% ± 1.10 | 92.31% |
| | Psychoticism | MLP | 86.83% ± 7.37 | 79.95% ± 3.07 | 87.50% |
| Living environment | Hostility | SMV | 83.94% ± 10.16 | 76.75% ± 6.93 | 88.14% |
| | Somatization | SVM | 84.14% ± 9.70 | 76.85% ± 7.98 | 87.50% |
| | Depression | MLP | 73.34% ± 3.67 | 71.04% ± 3.46 | 84.38% |
| | Obsession-Compulsion | RF | 95.35% ± 2.12 | 88.36% ± 0.78 | 89.55% |
| | Anxiety | SVM | 82.20% ± 10.67 | 75.35% ± 7.49 | 94.12% |
| | Interpersonal Sensitivity | GB | 94.44% ± 5.40 | 80.75% ± 1.74 | 83.82% |
| | Agoraphobia | MLP | 90.60% ± 7.83 | 81.22% ± 3.77 | 90.00% |
| | Paranoid Ideation | RF | 96.09% ± 1.98 | 85.63% ± 0.67 | 90.38% |
| | Psychoticism | RF | 96.50% ± 1.63 | 79.53% ± 1.35 | 92.86% |
| | Hostility | SVM | 89.25% ± 7.68 | 82.87% ± 5.57 | 91.53% |
| | Somatization | RF | 96.13% ± 1.38 | 87.72% ± 0.78 | 89.06% |
| | Depression | DT | 93.12% ± 2.29 | 78.73% ± 1.35 | 85.94% |
| | Obsession-Compulsion | RF | 97.64% ± 0.98 | 88.20% ± 0.71 | 88.06% |
| Economic status | Anxiety | RF | 97.17% ± 1.12 | 85.75% ± 0.51 | 85.29% |
| | Interpersonal Sensitivity | DT | 91.98% ± 3.01 | 81.55% ± 1.48 | 91.18% |
| | Agoraphobia | SVM | 88.01% ± 9.52 | 80.32% ± 8.16 | 94.00% |
| | Paranoid Ideation | RF | 95.32% ± 1.60 | 86.92% ± 0.89 | 96.15% |
| | Psychoticism | NB | 83.92% ± 4.15 | 82.60% ± 3.74 | 87.50% |
| | Hostility | DT | 95.74% ± 1.15 | 93.12% ± 0.35 | 96.61% |
| | Somatization | RF | 96.70% ± 0.15 | 96.15% ± 0.68 | 92.19% |
| | Depression | GB | 96.84% ± 0.79 | 95.37% ± 0.78 | 95.31% |
| Health impacts | Obsession-Compulsion | DT | 94.01% ± 0.61 | 93.51% ± 0.7 | 98.51% |
| | Anxiety | RF | 95.40% ± 0.16 | 95.31% ± 0.41 | 92.65% |
| | Interpersonal Sensitivity | DT | 95.68% ± 0.63 | 94.69% ± 0.7 | 92.65% |
| | Agoraphobia | NB | 92.79% ± 8.67 | 92.61% ± 8.91 | 98.00% |
| | Paranoid Ideation | DT | 92.30% ± 0.68 | 92.00% ± 1.13 | 100.0% |
| | Psychoticism | SVM | 87.33% ± 7.45 | 85.89% ± 7.39 | 91.07% |

A primary goal when training an ML model is to avoid overfitting or underfitting. To achieve this, we introduce the concept of the bias-variance trade-off (Belkin et al., 2019), which refers to balancing model complexity and its ability to generalize to unseen data (test set). This trade-off is key to optimizing model performance and robustness. Bias refers to a model's ability to capture patterns in the data, while variance reflects its sensitivity to small fluctuations during training. Accuracy can be used as a primary way to measure it, but more complex metrics should be used as those considering FP and FN. These kinds of metrics allow us to do a deeper interpretation of how the models perform. Like Table 6, we find Table 7 and Table 8 for Specificity and Sensitivity for the best models compiled regarding the Accuracy.

**Table 7.** Specificity for the ML classifiers using SA-45 labels.

| Training subset | Binary classifier | Train | Validation | Test |
|---|---|---|---|---|
| | Hostility | 94.35% ± 0.07 | 91.92% ± 0.50 | 96.67% |
| | Somatization | 99.19% ± 0.01 | 92.69% ± 0.20 | 93.75% |
| | Depression | 97.81% ± 0.02 | 94.09% ± 0.14 | 93.75% |
| Demographic characteristics | Obsession-Compulsion | 97.93% ± 0.01 | 94.09% ± 0.44 | 97.06% |
| | Anxiety | 98.10% ± 0.04 | 95.86% ± 0.08 | 100.0% |
| | Interpersonal Sensitivity | 95.56% ± 0.05 | 90.28% ± 0.45 | 97.06% |
| | Agoraphobia | 96.41% ± 0.02 | 90.09% ± 0.50 | 92.00% |
| | Paranoid Ideation | 97.82% ± 0.01 | 90.34% ± 0.50 | 88.46% |
| | Psychoticism | 96.89% ± 0.02 | 82.98% ± 0.76 | 92.86% |
| Living environment | Hostility | 100.0% ± 0.00 | 99.26% ± 0.02 | 96.67% |
| | Somatization | 99.41% ± 0.00 | 97.46% ± 0.10 | 93.75% |
| | Depression | 88.41% ± 1.24 | 81.01% ± 3.88 | 84.38% |
| | Obsession-Compulsion | 98.30% ± 0.01 | 92.49% ± 0.17 | 91.18% |
| | Anxiety | 98.68% ± 0.01 | 95.54% ± 0.15 | 97.06% |
| | Interpersonal Sensitivity | 98.11% ± 0.02 | 85.69% ± 0.59 | 94.12% |
| | Agoraphobia | 94.96% ± 0.02 | 88.03% ± 0.84 | 92.00% |
| | Paranoid Ideation | 100.0% ± 0.00 | 87.03% ± 0.63 | 88.46% |
| | Psychoticism | 98.00% ± 0.03 | 83.93% ± 0.58 | 96.43% |
| Economic status | Hostility | 95.30% ± 0.53 | 95.29% ± 2.11 | 91.44% |
| | Somatization | 95.24% ± 0.39 | 95.25% ± 1.57 | 87.50% |
| | Depression | 98.41% ± 0.20 | 98.43% ± 0.79 | 84.38% |
| | Obsession-Compulsion | 95.52% ± 0.75 | 95.54% ± 3.01 | 87.97% |
| | Anxiety | 92.59% ± 0.65 | 92.59% ± 2.62 | 85.29% |
| | Interpersonal Sensitivity | 91.85% ± 0.56 | 91.85% ± 2.22 | 94.12% |
| | Agoraphobia | 96.46% ± 0.87 | 96.39% ± 3.59 | 94.00% |
| | Paranoid Ideation | 94.71% ± 0.57 | 94.74% ± 2.30 | 96.15% |
| | Psychoticism | 93.75% ± 0.66 | 93.74% ± 2.67 | 89.29% |
| Health impacts | Hostility | 100.0% ± 0.00 | 100.0% ± 0.00 | 100.0% |
| | Somatization | 100.0% ± 0.00 | 100.0% ± 0.00 | 100.0% |
| | Depression | 100.0% ± 0.00 | 100.0% ± 0.00 | 100.0% |
| | Obsession-Compulsion | 99.25% ± 0.00 | 99.29% ± 0.02 | 100.0% |
| | Anxiety | 99.26% ± 0.00 | 99.33% ± 0.02 | 94.12% |
| | Interpersonal Sensitivity | 99.26% ± 0.00 | 99.33% ± 0.02 | 97.06% |
| | Agoraphobia | 100.0% ± 0.00 | 100.0% ± 0.00 | 100.0% |
| | Paranoid Ideation | 100.0% ± 0.00 | 100.0% ± 0.00 | 100.0% |
| | Psychoticism | 100.0% ± 0.00 | 100.0% ± 0.00 | 100.0% |

**Table 8.** Sensitivity for the ML classifiers using SA-45 labels.

| Training subset | Binary classifier | Train | Validation | Test |
|---|---|---|---|---|
| Demographic characteristics | Hostility | 88.61% ± 0.19 | 81.76% ± 0.84 | 65.52% |
| | Somatization | 78.19% ± 0.09 | 72.03% ± 1.01 | 81.25% |
| | Depression | 97.81% ± 0.02 | 94.09% ± 0.14 | 81.25% |
| | Obsession-Compulsion | 83.02% ± 0.06 | 75.55% ± 0.79 | 81.82% |
| | Anxiety | 76.77% ± 0.05 | 98.10% ± 0.04 | 79.41% |
| | Interpersonal Sensitivity | 83.40% ± 0.16 | 67.63% ± 0.22 | 73.53% |
| | Agoraphobia | 91.67% ± 0.01 | 96.41% ± 0.02 | 90.00% |
| | Paranoid Ideation | 89.41% ± 0.05 | 78.61% ± 1.77 | 96.15% |
| | Psychoticism | 96.26% ± 0.05 | 87.44% ± 0.53 | 82.14% |
| Living environment | Hostility | 82.28% ± 0.04 | 78.52% ± 1.22 | 79.31% |
| | Somatization | 83.13% ± 0.03 | 77.86% ± 0.43 | 81.25% |
| | Depression | 74.38% ± 2.72 | 69.23% ± 2.71 | 84.38% |
| | Obsession-Compulsion | 88.25% ± 0.05 | 81.33% ± 0.54 | 87.88% |
| | Anxiety | 83.25% ± 0.04 | 76.32% ± 0.56 | 91.18% |
| | Interpersonal Sensitivity | 88.26% ± 0.02 | 75.43% ± 0.78 | 81.25% |
| | Agoraphobia | 86.85% ± 0.06 | 78.68% ± 0.18 | 88.00% |
| | Paranoid Ideation | 95.13% ± 0.03 | 78.07% ± 0.24 | 92.31% |
| | Psychoticism | 94.37% ± 0.04 | 80.58% ± 0.21 | 89.29% |
| | Hostility | 95.30% ± 0.53 | 95.29% ± 2.11 | 91.44% |
| | Somatization | 95.24% ± 0.39 | 95.25% ± 1.57 | 87.50% |
| | Depression | 98.41% ± 0.20 | 98.43% ± 0.79 | 84.38% |
| | Obsession-Compulsion | 95.52% ± 0.75 | 95.54% ± 3.01 | 87.97% |
| Economic status | Anxiety | 92.59% ± 0.65 | 92.59% ± 2.62 | 85.29% |
| | Interpersonal Sensitivity | 91.85% ± 0.56 | 91.85% ± 2.22 | 94.12% |
| | Agoraphobia | 96.46% ± 0.87 | 96.39% ± 3.59 | 94.00% |
| | Paranoid Ideation | 94.71% ± 0.57 | 94.74% ± 2.30 | 96.15% |
| | Psychoticism | 93.75% ± 0.66 | 93.74% ± 2.67 | 89.29% |
| | Hostility | 94.02% ± 0.00 | 91.05% ± 0.15 | 93.10% |
| | Somatization | 93.66% ± 0.01 | 93.71% ± 0.22 | 84.38% |
| | Depression | 94.44% ± 0.00 | 92.00% ± 0.29 | 90.63% |
| Health impacts | Obsession-Compulsion | 90.29% ± 0.02 | 86.31% ± 0.65 | 96.97% |
| | Anxiety | 91.87% ± 0.00 | 91.94% ± 0.06 | 91.18% |
| | Interpersonal Sensitivity | 93.28% ± 0.02 | 90.19% ± 0.28 | 88.24% |
| | Agoraphobia | 92.88% ± 0.02 | 92.22% ± 0.22 | 96.00% |
| | Paranoid Ideation | 85.61% ± 0.03 | 86.18% ± 0.64 | 100.0% |
| | Psychoticism | 83.98% ± 0.08 | 83.92% ± 0.92 | 82.14% |

**K-Means labels.** The second use case corresponds to the classifiers trained with the labels obtained after applying K-Means. In this case, we have three diagnostic classes, less vulnerable, vulnerable and highly vulnerable against control individuals. Using this data, we apply various binary classification approaches to assess an individual's level of psychological vulnerability or determine whether they do not experience any psychological conditions.

This results in the findings presented in Table 9, where the accuracy of each of the three classifiers using the four subsets is reported for the training, validation, and test subsets. The values are expressed as the mean and standard deviation, reflecting the use of k-fold validation.

**Table 9.** Accuracy for the ML classifiers.

| Training subset | Binary classifier | Best ML model | Train | Validation | Test |
|---|---|---|---|---|---|
| | Highly Vulnerable | MLP | 97.67% ± 1.56 | 90.96% ± 2.79 | 95.59% |
| Demographic characteristics | Less Vulnerable | SVM | 91.01% ± 0.77 | 84.86% ± 5.44 | 90.24% |
| | Vulnerable | RF | 97.46% ± 0.46 | 93.11% ± 2.61 | 96.38% |
| | Highly Vulnerable | MLP | 96.74% ± 3.85 | 92.10% ± 3.87 | 96.10% |
| Living environment | Less Vulnerable | RF | 96.00% ± 0.77 | 83.12% ± 3.98 | 85.90% |
| | Vulnerable | MLP | 99.07% ± 1.20 | 94.05% ± 2.17 | 94.30% |
| | Highly Vulnerable | HGB | 96.85% ± 0.79 | 91.56% ± 2.78 | 95.74% |
| Economic status | Less Vulnerable | SVM | 87.43% ± 1.10 | 78.47% ± 4.45 | 83.15% |
| | Vulnerable | RF | 98.32% ± 0.56 | 91.00% ± 2.68 | 97.83% |
| | Highly Vulnerable | SVM | 98.62% ± 0.15 | 97.41% ± 1.43 | 98.53% |
| Health impacts | Less Vulnerable | SVM | 97.43% ± 0.62 | 93.68% ± 3.16 | 93.75% |
| | Vulnerable | DT | 96.99% ± 0.53 | 96.65% ± 2.14 | 97.50% |

By looking at the results above, we can check if there is any overfitting or underfitting issue by using the concept of bias-variance trade off. As in the case of the SA-45 classifiers accuracy serves as a primary metric for evaluation, but more advanced metrics using FP and FN, provide a more comprehensive assessment of model performance. Similarly, we have obtained the values for Specificity and Sensitivity compiled in Tables 10 and 11.

**Table 10.** Specificity for the ML classifiers.

| Training subset | Binary classifier | Train | Validation | Test |
|---|---|---|---|---|
| Demographic characteristics | Highly Vulnerable | 97.67% ± 1.56 | 90.96% ± 2.78 | 95.59% |
| | Less Vulnerable | 91.01% ± 0.77 | 84.88% ± 5.41 | 90.24% |
| | Vulnerable | 97.46% ± 0.46 | 93.11% ± 2.62 | 96.38% |
| Living environment | Highly Vulnerable | 96.74% ± 3.85 | 92.10% ± 3.87 | 96.10% |
| | Less Vulnerable | 96.00% ± 0.77 | 83.12% ± 4.02 | 85.09% |
| | Vulnerable | 99.07% ± 1.20 | 94.55% ± 2.63 | 94.30% |
| Economic status | Highly Vulnerable | 96.85% ± 0.79 | 91.56% ± 2.78 | 95.76% |
| | Less Vulnerable | 87.43% ± 1.10 | 78.49% ± 4.42 | 83.16% |
| | Vulnerable | 98.32% ± 0.56 | 91.00% ± 2.68 | 97.83% |
| Health impacts | Highly Vulnerable | 98.62% ± 0.14 | 97.41% ± 1.43 | 98.53% |
| | Less Vulnerable | 97.43% ± 0.62 | 93.68% ± 3.16 | 93.75% |
| | Vulnerable | 96.99% ± 0.53 | 96.65% ± 2.14 | 97.50% |

4**Table 11.** Sensitivity for the ML classifiers.

| Training subset | Binary classifier | Train | Validation | Test |
|---|---|---|---|---|
| Demographic characteristics | Highly Vulnerable | 97.67% ± 1.56 | 90.96% ± 2.78 | 95.59% |
| | Less Vulnerable | 91.01% ± 0.77 | 84.88% ± 5.41 | 90.24% |
| | Vulnerable | 97.46% ± 0.46 | 93.11% ± 2.62 | 96.38% |
| Living environment | Highly Vulnerable | 96.74% ± 3.85 | 92.10% ± 3.87 | 96.10% |
| | Less Vulnerable | 96.00% ± 0.77 | 83.12% ± 4.02 | 85.09% |
| | Vulnerable | 99.07% ± 1.20 | 94.05% ± 2.17 | 94.30% |
| Economic status | Highly Vulnerable | 96.85% ± 0.79 | 91.56% ± 2.78 | 95.76% |
| | Less Vulnerable | 87.44% ± 1.10 | 78.49% ± 4.42 | 83.16% |
| | Vulnerable | 98.32% ± 0.56 | 91.00% ± 2.68 | 97.83% |
| Health impacts | Highly Vulnerable | 98.62% ± 0.14 | 97.41% ± 1.43 | 98.53% |
| | Less Vulnerable | 97.43% ± 0.62 | 93.68% ± 3.16 | 93.75% |
| | Vulnerable | 96.99% ± 0.53 | 96.65% ± 2.14 | 97.50% |

### 4.3 Influential features

Finally, after achieving a classification model with reasonable accuracy, the next step is to identify which features significantly impact its predictions. This can be done by using feature extraction techniques such as SHAP. We have applied it to each of the three use cases.

For the case of labelled individuals using only the SA-45 test, we have obtained nine binary classifiers each one for the different psychological conditions diagnosed by the test. In Table 12, we compiled all the information related to these conditions and show the 3 most influential features for each of the subsets: personal data, living conditions, economic features and general health. To choose which are the features that really have a notable influence on the classifiers, we have bolded those having values over 0.75.

**Table 12.** SHAP values for SA-45 labels dataset.

| Binary classifier | Variables subset | Shape value | Top 3 |
|---|---|---|---|
| Hostility | Personal data | Marital status | 0.2940 |
| | | Living arrangement pre-quarantine | 0.2614 |
|  |  | Highest education completed | 0.2554 |
|  | Living conditions | Living Space During Confinement | 0.2508 |
|  |  | Useful housing m² | 0.2109 |
|  |  | Weekly outings quarantine | 0.1828 |
|  | Economic features | work situation pre-quarantine | 0.1849 |
|  |  | situation during the quarantine | 0.1505 |
|  |  | Primary earner's occupation | 0.1350 |
|  | Health impacts | Health condition | 0.2887 |
|  |  | Psychological therapy ongoing | 0.2445 |
|  |  | Family COVID cohabitation | 0.2089 |
| Somatization | Personal data | Current documentation status | 0.3607 |
|  |  | Contacts during quarantine | 0.2852 |
|  |  | Marital status | 0.1485 |
|  | Living conditions | Useful housing m² | 0.2303 |
|  |  | Living space during confinement | 0.1852 |
|  |  | Weekly outings quarantine | 0.1577 |
|  | Economic features | situation during the quarantine | 0.2602 |
|  |  | Pre-quarantine household income | 0.1975 |
|  |  | Primary earner's occupation | 0.0738 |
|  | Health impacts | Health condition | 0.3456 |
|  |  | COVID-19 diagnosis | 0.2162 |
|  |  | Family COVID cohabitation | 0.2064 |
| Depression | Personal data | Contacts during quarantine. | 0.2531 |
|  |  | Marital status. | 0.1516 |
|  |  | Highest education completed. | 0.1336 |
|  | Living conditions | Useful housing m² | 0.2512 |
|  |  | Weekly outings quarantine | 0.2217 |
|  |  | Living space during confinement | 0.2153 |
|  | Economic features | Situation during the quarantine | 0.3220 |
| | | Pre-quarantine household income. | 0.2581 |
| | | work situation pre-quarantine | 0.1058 |
| | Health impacts | **Health condition** | **11.061** |
| | | **Psychological therapy ongoing** | **10.238** |
| | | **Family COVID cohabitation** | **8.4911** |
| Obsession-Compulsion | Personal data | Contacts during quarantine | 0.2269 |
| | | Marital status | 0.1924 |
| | | Living arrangement pre-quarantine | 0.1398 |
| | Living conditions | Useful housing m² | 0.2341 |
| | | Weekly outings quarantine | 0.1433 |
| | | Living space during confinement | 0.1421 |
| | Economic features | Household income quarantine | 0.1814 |
| | | situation during the quarantine | 0.1533 |
| | | work situation pre-quarantine | 0.1133 |
| | Health impacts | Family COVID cohabitation | 0.2674 |
| | | Health condition | 0.2600 |
| | | Psychological therapy ongoing | 0.2442 |
| Anxiety | Personal data | Contacts during quarantine | 0.2754 |
| | | Marital status | 0.1430 |
| | | Usual living arrangement | 0.1377 |
| | Living conditions | Useful housing m² | 0.2122 |
| | | Living space during confinement | 0.1762 |
| | | Weekly outings quarantine | 0.1347 |
| | Economic features | Work situation during the quarantine | 0.1845 |
| | | Post-crisis employment outlook | 0.1510 |
| | | Household income quarantine | 0.1127 |
| | Health impacts | Health condition | 0.3434 |
| | | Family COVID cohabitation | 0.1814 |
| | | Past suicide attempts | 0.1764 |
| Interpersonal Sensitivity | Personal data | Marital status | 0.3405 |
| | | Contacts pre-quarantine | 0.2825 |
| | | Contacts during quarantine | 0.2002 |
| | Living conditions | Useful housing m² | 0.2176 |
| | | Living space during confinement | 0.1957 |
| | | Residence quality quarantine | 0.1285 |
| | Economic features | situation during the quarantine | 0.2575 |
| | | Pre-quarantine household income | 0.2152 |
| | | Primary earner's occupation | 0.2082 |
| | Health impacts | Psychological therapy ongoing | 0.3331 |
| | | Health condition | 0.3315 |
| | | Family COVID cohabitation | 0.2888 |
| Agoraphobia | Personal data | Contacts during quarantine | 0.2381 |
| | | Marital status | 0.1917 |
| | | Highest education completed | 0.1398 |
| | Living conditions | **Useful housing m²** | **2.2129** |
| | | **Weekly outings quarantine** | **2.0153** |
| | | **Residence quality quarantine** | **1.3185** |
| | Economic features | Work situation during the quarantine | 0.2317 |
| | | Work situation pre-quarantine | 0.1789 |
| | | Pre-quarantine household income | 0.1235 |
| | Health impacts | Health condition | 0.2748 |
| | | COVID-19 diagnosis | 0.1995 |
| | | Family COVID cohabitation | 0.1177 |
| Paranoid Ideation | Personal data | Contacts during quarantine. | 0.3230 |
| | | Highest education completed | 0.2824 |
| | | Marital status | 0.2279 |
| | Living conditions | Weekly outings quarantine | 0.2435 |
| | | Living space during confinement | 0.2410 |
| | | Useful housing m² | 0.2003 |
| | Economic features | situation during the quarantine | 0.2404 |
| | | Pre-quarantine household income | 0.1097 |
| | | Primary earner's occupation | 0.0917 |
| | Health impacts | COVID-19 diagnosis | 0.4156 |
| | | Family COVID cohabitation | 0.3724 |
| | | Health condition | 0.3413 |
| Psychoticism | Personal data | Current documentation status | 0.3863 |
| | | Contacts during quarantine | 0.2080 |
| | | Contacts pre-quarantine | 0.1733 |
| | Living conditions | Useful housing m² | 0.2637 |
| | | Weekly outings quarantine | 0.2369 |
| | | Living space during confinement | 0.1903 |
| | Economic features | Household Conditions During Confinement | 0.2234 |
| | | Pre-quarantine household income | 0.1882 |
| | | Situation during the quarantine | 0.1547 |
| | Health impacts | Chronic Illness | 0.2622 |
| | | Health condition | 0.2492 |
| | | Psychological therapy ongoing | 0.2478 |

To an end and for a better understand of the impact of these features on the classifiers, we refer to Table 13, which presents the most frequently occurring values for each the notable influential features

**Table 13.** Top distributed values for each of the most influential feature.

| Feature | Total points |
|---|---|
| Health condition | Any condition: 64.55% |
| Psychological therapy ongoing | No: 65.18% |
| Family COVID cohabitation | No: 54.43% |
| | Does not proceed: 40.5% |
| Useful housing m² | 91 to 105 m²: 18.55% |
| | More than 180 m²: 16.94% |
|  | 30 to 45 m²: 16.94% |
|  | 106 to 120 m²: 14.52% |
| Weekly outings quarantine | Each 10 ten days approximately: 31.45% |
|  | I have not gone out: 27.42% |
| Residence quality quarantine | Country house: 80.65% |

Next, we are obtaining the most influential features for the case of vulnerability level classifiers. As before, we are using SHAP values and applied to the three classifiers trained with the four subsets. In Table 14, we compiled all the information related to the three levels of vulnerability showing which are the features with higher SHAP values. Again, the features having a notable influence are bolded.

Table 14. SHAP values for level of vulnerability classifiers.

| Binary classifier | Features subset | Top 3 Features | SHAP Values |
|---|---|---|---|
| Highly Vulnerable | Demographic characteristics | Living arrangement in quarantine | 0.6167 |
|  |  | Adults in Care in quarantine | 0.4671 |
|  |  | Living arrangement pre-quarantine | 0.4444 |
|  | Living environment | Usable living space | 0.4503 |
|  |  | Household conditions during Confinement | 0.3858 |
|  |  | Bedroom cohabitants count | 0.3800 |
|  | Economic status | **Household conditions during confinement** | **2.4960** |
|  |  | **Situation during the quarantine** | **1.4991** |
|  |  | **Post-crisis employment outlook** | **1.2139** |
|  | Health impacts | Isolation during covid diagnosis | 0.7197 |
|  |  | Past suicide attempts | 0.6840 |
|  |  | In need of assistance | 0.6654 |
| Less Vulnerable | Demographic characteristics | Nationality | 0.6445 |
|  |  | Contacts pre-quarantine | 0.4892 |
|  |  | Current documentation status | 0.4791 |
|  | Living environment | Residence quality quarantine | 0.2897 |
| | | Weekly outings quarantine | 0.2083 |
| | | Bedroom cohabitants count | 0.1948 |
| | Economic status | Household conditions during confinement | 0.4519 |
| | | Situation during the quarantine | 0.4031 |
| | | Highest economic contributor pre-quarantine | 0.3327 |
| | Health impacts | Family member COVID severity | 0.5607 |
| | | Past suicide attempts | 0.5348 |
| | | Isolation during covid diagnosis | 0.5286 |
| Vulnerable | Demographic characteristics | Marital status | 0.2669 |
| | | Highest education completed | 0.1905 |
| | | Contacts during quarantine | 0.1857 |
| | Living environment | Cohabitants count | 0.6178 |
| | | Weekly outings quarantine | 0.3895 |
| | | Living Space in quarantine | 0.3716 |
| | Economic status | Household conditions during confinement | 0.4099 |
| | | Household income quarantine | 0.2742 |
| | | Occupation highest economic contributor pre-quarantine | 0.2376 |
| | Health impacts | Disability degree | 0.6614 |
| | | Past suicide attempts | 0.6179 |
| | | Covid severity | 0.5711 |

Finally, we compile in Table 15 the most common values for each of the key influential features regarding the previous table.

Table 15. Distribution of most influential features

| Feature | Features' values with a great percentage |
|---|---|
| Household Conditions During Confinement | Enough money to support themselves (83.66%) |
| Situation during the quarantine | Telework employee (32.48%) <br> Record of Temporary Employment Regulation (16.45%) |
| Post-crisis employment outlook | Presential employee (40.55%) |

## 5   Discussion

As we have said before the aim of this work is to find which features from different subsets influence more in the psychological conditions of a population that was exposed to the COVID quarantine in Spain. In this way, two use cases have been raised. The first one only uses the SA-45 where we have implemented different diagnostic classifiers for nine psychological conditions. The second case, first, has clustered individuals regarding its psychological vulnerability degree and then has obtained different classifiers for assigning these levels of vulnerability. All the models have been trained with four different subsets whose features have been grouped depending on their nature.

For the first use case, we first obtained the accuracy metric for the different diagnostic models that depend on the nine psychological conditions. This information which is compiled in Table 6 shows the following. The evaluated ML models demonstrated satisfactory performance (high bias), with accuracy rates exceeding 80% and many surpassing 90%. Variance was also acceptable as it remained within a 10% difference. When discussing the most effective ML models, RF, DT, and SVM stand out, achieving the best performance in 11, 9, and 8 use cases, respectively.

We also have measured the performance of the models using metrics such as specificity and sensitivity. These metrics are essential for understanding the types of errors a model makes, especially in contexts where FP or FN carry significant consequences. Tables 7 and 8 provide sensitivity and specificity results comparable to those in Table 6. Since the best-performing model remains consistent across these evaluations, its column has been excluded from these tables.

In the case of Table 7, most models achieve accuracy rates exceeding 80%, with many surpassing 90%. However, performance varies in specific cases. For the subset of demographic characteristics models are good diagnosing each psychological condition. Within the living environment subset, depression has more problems to be diagnosed. The economic status subset reveals good diagnosis for each psychological condition. Also, the health impacts subset demonstrates strong performance across all

29psychological conditions, with no significant issues detected. These results help to evaluate whether the models struggle to accurately identify healthy individuals, potentially leading to misdiagnoses of psychological conditions. Such errors could result in unnecessary treatments, with both economic and psychological repercussions.

In the case of the metrics in Table 8 hold particular significance due to their direct impact on mental health outcomes. Low values suggest the risk of individuals with psychological conditions being misdiagnosed as healthy, which may put their well-being and access to appropriate care at risk. Among the subsets analysed, the health impacts subset shows the strongest performance, consistent with the high sensitivity values observed. For the demographic characteristics, the models have accurate values in many cases but has bad performance in the case of anxiety and interpersonal sensitivity. For the living environment performance, most of the classifiers range from 80% to around 90% and only in the case of hostility the values are under 80%. Finally, the results for economic status are good for all the psychological issues.

For the second use case, we have obtained the same metrics as in the first one. The main difference is that we have obtained three diagnostic models that discriminate between levels of vulnerability and controls. What remains the same is that these models were trained using the same four subsets. The first studied metric is the accuracy compiled in Table 9. The evaluated ML models exhibited strong performance with high bias, achieving accuracy rates above 90%, except in the case of less vulnerable individuals with living environment and economic features which is around 85% and 83% respectively. Additionally, variance remained within an acceptable range, with differences staying under 10% in most of the cases. By analysing the best performing, SVM are the best in 4 cases.

Other metrics such as specificity and sensitivity are compiled in Table 10 and 11 following the same structure as Table 9 except for the best performing model with remains the same for the three tables. In both cases, the metrics obtain similar values than in Table 9 with values over 90% in test and low variances. Again, the worst result is for the same case, the classifier for less vulnerable individuals using the living environment and the economic status subsets that obtains values a bit higher than 83%. So, we can conclude that the models only have some problems in diagnosing individuals that are less vulnerable.

Once the models have been trained, we aim to identify the features that have the greatest influence on the diagnosis of each psychological condition and the level of vulnerability. Since our features are grouped into four subsets, we analyse each group individually. To this end, we calculated the SHAP values in both uses cases for the best-performing classifiers, as indicated in Table 6 and Table 9. SHAP values close to 1.0 indicate strong influence, so we have considered features in the top quartile (greater than 0.75) as important.

Considering this criterion we have created Table 12 for the use case of SA-45 classifiers. Analysing the results, we can see the following interesting points. Depression is strongly influenced by the subset of health conditions standing out the general health condition, if the individual is receiving a psychological treatment and if there was cohabitation with people with COVID. In the case of agoraphobia, the subset of the living conditions has a big impact due to the size of the house, the weekly outgoings and the quality of the residence during the quarantine. For the remaining psychological conditions, no features exhibit a strong influence. To gain a deeper understanding of how these features influence the classifiers, we can refer to Table 13, which highlights the most frequently occurring values for each influential feature. What we can conclude with these values is the following. Having a health condition and not doing therapy influence depression. In the case of agoraphobia, making many or no outings is linked to agoraphobia. For this psychological, the fact of living in a country house is also a trigger.

Finally, we have done the same process related to the SHAP values for the second use case of diagnosing level of vulnerability. In this case and regarding Table 14, there is only a set of features that has a high influence in diagnosing highly vulnerable individuals. These features belong to the economic status subset and describe the household conditions, confinements situation during the quarantine and post-crisis employments outlook. Finally, we have obtained the distribution of the different values of this features which are compiled in Table 15. These findings suggest that the absence of economic problems is not a determining factor in experiencing a higher number of psychological conditions. Additionally, there is an observable influence on individuals who continued working remotely, were temporarily laid off, or were considering the possibility of returning to the office.

# 6   Conclusions and future works

The present study explores the impact of personal, home-environmental, economic, and health-related factors on psychological conditions experienced during the COVID-19 lockdown in Spain. By leveraging AI techniques, particularly both supervised and unsupervised models, we first develop diagnostic models and then identify the most influential features in making these diagnoses. The experiments consist of two use cases: one that relies solely on classical ML classifiers and another that first applies clustering to create groups, which are then used as input for a second set of ML classifiers. In the first approach, SA-45 test labels serve as class labels for different models such as SVM, MLP, or RF. The second approach applies K-Means clustering to analyze individuals' behavior based on the SA-45 test, using the clustering results (levels of vulnerability) to label individuals, who are then used to train the same set of classical ML classifiers. Finally, for both cases, we apply SHAP value techniques to determine the most influential features and assess how their values impact the diagnosis. The results demonstrate that diagnostic approaches can effectively use different sets of features to classify individuals depending on mental health conditions. Depression is strongly associated with health-related factors, while agoraphobia is mainly influenced by living conditions. Other psychological conditions showed no significant feature associations. SHAP analysis for diagnosing vulnerability revealed that economic factors, such as household conditions and post-crisis employment outlook, play a key role. Notably, economic stability did not necessarily correlate with fewer psychological conditions. These findings provide insights into the complex interplay between health, living conditions, and economic status in mental well-being during crises.

Despite the promising results, this study presents certain limitations. First, the dataset was collected through self-reported questionnaires, which may introduce biases related to subjective perception and recall accuracy. Second, the study focuses exclusively on the Spanish population during the 2020 lockdown, limiting the generalizability of results to other cultural and policy contexts.

For future work, several proposals could be developed. Applying the same methodology to datasets from other countries could validate and extend the findings, allowing comparisons. Developing predictive models for early detection and intervention could assist policymakers in designing targeted mental health support mechanisms for at-risk groups. Studying the long-term psychological consequences of lockdown measures could offer valuable information on resilience factors and potential recovery patterns.

**Funding Declaration**

No funding was received for conducting this study.

**Competing Interest**

The author(s) declare(s) that they have no competing interests.

33Jojoa, M., Garcia-Zapirain, B., Gonzalez, M. J., Perez-Villa, B., Urizar, E., Ponce, S., & Tobar-Blandon, M. F. (2022). Analysis of the Effects of Lockdown on Staff and Students at Universities in Spain and Colombia Using Natural Language Processing Techniques. *International Journal of Environmental Research and Public Health*, *19*(9), 5705. https://doi.org/10.3390/ijerph19095705

Kirkbride, J. B., Anglin, D. M., Colman, I., Dykxhoorn, J., Jones, P. B., Patalay, P., Pitman, A., Soneson, E., Steare, T., Wright, T., & Griffiths, S. L. (2024). The social determinants of mental health and disorder: evidence, prevention and recommendations. *World Psychiatry*, *23*(1), 58–90. https://doi.org/10.1002/wps.21160

Kohonen, T. (1982). Self-organized formation of topologically correct feature maps. *Biological Cybernetics*, *43*(1), 59–69. https://doi.org/10.1007/BF00337288

Kohonen, T. (1998). The self-organizing map. *Neurocomputing*, *21*(1), 1–6. https://doi.org/https://doi.org/10.1016/S0925-2312(98)00030-7

Luijten, M. A. J., van Muilekom, M. M., Teela, L., van Oers, H. A., Terwee, C. B., Zijlmans, J., Klaufus, L., Popma, A., Oostrom, K. J., Polderman, T. J. C., & Haverman, L. (2020). The impact of lockdown during the COVID-19 pandemic on mental and social health of children and adolescents. *MedRxiv*. https://doi.org/10.1101/2020.11.02.20224667

Lundberg, S. M., Allen, P. G., & Lee, S.-I. (n.d.). *A Unified Approach to Interpreting Model Predictions*. https://github.com/slundberg/shap

Mccullagh, P. (2008). Sampling bias and logistic models. In *J. R. Statist. Soc. B* (Vol. 70). https://academic.oup.com/jrsssb/article/70/4/643/7109508

Mellor-Marsá, B., Guitián, A., García-Tejedor, Á. J., & Nogales, A. (2024). *LOCKED: A Dataset of Sociodemographic, Economic, Health and Living Features to Assess Mental Health Impact of the Spanish Lockdown during COVID-19*. https://doi.org/10.20944/preprints202412.0310.v1

Mitchell, H. B. (n.d.). Bayesian Decision Theory. In *Multi-Sensor Data Fusion* (pp. 201–219). Springer Berlin Heidelberg. https://doi.org/10.1007/978-3-540-71559-7_12

Montero-Marin, J., Hinze, V., Mansfield, K., Slaghekke, Y., Blakemore, S.-J., Byford, S., Dalgleish, T., Greenberg, M. T., Viner, R. M., Ukoumunne, O. C., Ford, T., & Kuyken, W. (2023). Young People's Mental Health Changes, Risk, and Resilience During the COVID-19 Pandemic. *JAMA Network Open*, *6*(9), e2335016. https://doi.org/10.1001/jamanetworkopen.2023.35016

Mora, T., Fichera, E., & Lopez-Valcarcel, B. G. (2023). How has the strict lockdown during the SARS-COV-2 outbreak changed the diet of Spaniards? *SSM - Population Health*, *24*, 101512. https://doi.org/10.1016/j.ssmph.2023.101512

Morgül, E., Kallitsoglou, A., & Essau, C. A. E. (2020). Psychological effects of the COVID-19 lockdown on children and families in the UK. *Revista de Psicolog{\'i}a Cl{\'i}nica Con Niños y Adolescentes*, *7*(3), 42–48.

Ntakolia, C., Priftis, D., Charakopoulou-Travlou, M., Rannou, I., Magklara, K., Giannopoulou, I., Kotsis, K., Serdari, A., Tsalamanios, E., Grigoriadou, A., Ladopoulou, K., Koullourou, I., Sadeghi, N., OCallaghan, G., & Lazaratou, E. (2022). An Explainable Machine Learning Approach for COVID-19's Impact on

**Appendix A: Personal questionnaire.**

1. Please indicate the gender you identify with.
2. Date of birth (to select your year, click on the arrow next to the date and then scroll down the sidebar).
3. Indicate your marital status.
4. Please specify your place of birth.
5. What is the current status of your documentation?
6. Do you have European nationality?
7. What is the highest level of education you have completed?
8. Please provide details of your studies, if applicable.
9. Please indicate your usual living arrangements during the 12 months BEFORE the Lockdown.
10. Please indicate your usual living arrangements DURING the lockdown.
11. Please indicate the number of MINORS in your care DURING the quarantine.
12. Please indicate the number of dependents over 18 years old in your care (including elderly individuals, and people with disabilities) DURING the quarantine.
13. Please indicate how many people you usually had face-to-face contact with on a normal day BEFORE the quarantine (including at home, work, and socially).
14. Please indicate how many people you usually had face-to-face contact with on a normal day DURING the quarantine (including at home, work, and socially).
15. Please indicate your Postal Code during the lockdown.
16. Please specify the type of space you were living in DURING the lockdown.
17. Please indicate if you own your home or if you are renting.
18. How many usable square meters (that you can walk on) did your residence have DURING the quarantine?
19. Would you say your living space DURING the quarantine had adequate ventilation?
20. Would you say your living space DURING the quarantine had sufficient natural light?
21. Please indicate if these elements were present in your living space DURING the Quarantine.
22. Considering your place of residence DURING the lockdown, please indicate the number of rooms (not counting the bathroom and kitchen).
23. Considering your place of residence DURING the lockdown, please indicate the number of people you lived with.
24. Considering your place of residence DURING the lockdown, please indicate the number of people in your bedroom, excluding yourself.
25. Employment status BEFORE (in the 12 months before the start of the quarantine or most of the time).

26. Please provide details of the type of job during that period, if applicable.
27. What was the duration of your workday BEFORE (in the 12 months before the start of the quarantine or most of that time)?
28. What was your employment status DURING the quarantine?
29. What do you consider could be your employment or academic status AFTER the crisis caused by the coronavirus (in the 12 months following the end of the quarantine or most of that time)?
30. In the 12 months BEFORE the lockdown, what was the main occupation of the PERSON who contributed the most economic support to the HOUSEHOLD?
31. What was the approximate level of regular MONTHLY net income in your HOUSE-HOLD (unit where expenses are shared: individual, couple, family) BEFORE the quarantine (in the 12 months before the start of the quarantine or most of that time)?
32. What was the approximate level of regular MONTHLY net income in your HOUSE-HOLD (unit where expenses are shared: individual, couple, family) DURING the quarantine?
33. DURING the confinement in your home, did you experience any significant changes in your financial situation?
34. Suppose you (and your spouse or partner) convert all your funds in current and/or savings accounts, stock market investments, bonds, real estate, and sell your house, vehicles, and all your valuable items into money. Then, suppose you use the money from all these transactions to pay your mortgage and other credits, loans, debts, and credit cards. Would you still have money left after paying all your debts, or would you still owe money (a rough estimate is sufficient)?
35. Please indicate if you currently have any of the following health conditions.
36. If you have ever received a psychiatric diagnosis, please indicate which one.
37. Are you currently receiving psychiatric treatment or medication?
38. Are you currently receiving psychological treatment or therapy?
39. If you have consumed substances weekly in the last 6 months, please specify which ones.
40. Please indicate if you need help with daily self-care tasks such as shopping, household chores, bathing, grooming, cooking, managing money, etc.
41. Please indicate, if applicable, the degree of disability according to your certificate % recognized (0 if none).
42. Please indicate if you have ever attempted suicide.
43. Please indicate if you have been diagnosed with a coronavirus infection.
44. If you have had a coronavirus diagnosis, how would you rate the severity of the illness?
45. In this case, did you remain isolated inside your home (without leaving a room and without company during the duration of symptoms and 15 more

    days)?
46. If any member of the family unit has been diagnosed with a coronavirus infection, please indicate who it is (check more than one option if applicable).
47. If so, please assess the severity of the disease (consider the most severe case if there are multiple cases).
48. Has any member of your family diagnosed with a coronavirus infection lived in your home during their illness?
49. How many times a week did you leave the house during the quarantine?
50. For what reasons did you leave the house during the quarantine (select as many options as needed).
51. Do you consider that the measures taken to prevent the pandemic's progression are adequate and fair?

**Appendix B: SA-45 Test.**

1. The idea that another person can control your thoughts
2. Believing that most of your problems are someone else's fault
3. Feeling scared in open spaces or on the street
4. Hearing voices that other people do not hear
5. The idea that most people cannot be trusted
6. Feeling sudden and irrational fear
7. Outbursts of anger or rage that you cannot control
8. Fear of going out alone
9. Feeling lonely
10. Feeling sad
11. Losing interest in things
12. Feeling nervous or very anxious
13. Believing that others are aware of your thoughts
14. Feeling that others do not understand or listen to you
15. Having the impression that people are unfriendly or that you are disliked
16. Having to do things very slowly to be sure you are doing them right
17. Feeling inferior to others
18. Muscle pain
19. The feeling that others are watching or talking about you
20. Having to check everything you do repeatedly
21. Having difficulty making decisions
22. Feeling afraid to travel by bus, subway, or train
23. Feeling hot or cold suddenly
24. Having to avoid certain places or situations because they scare you
25. Mind going blank
26. Numbness or tingling in any part of your body
27. Feeling hopeless about the future
28. Having difficulty concentrating
29. Feeling weak in any part of your body
30. Feeling worried, tense, or agitated
31. Heaviness in arms or legs
32. Feeling uncomfortable when people look at you or talk about you
33. Having thoughts that are not yours
34. Feeling the urge to hit, hurt, or harm someone
35. Feeling like breaking something
36. Feeling very shy among other people
37. Feeling scared or anxious in crowded places (like a cinema or supermarket)
38. Panic or terror attacks
39. Having frequent arguments
40. Feeling that others do not adequately recognize your achievements
41. Feeling restless or uneasy
42. The feeling of being useless or worthless
43. Shouting or throwing things

44. The impression that people would try to take advantage of you if they could
45. The idea that you should be punished for your sins

**Appendix C: Distribution of initial features among the four subsets**

Demographic characteristics:
- Gender
- Age
- Nationality
- Marital Status
- Current Documentation Status
- Education Level
- Family Structure before COVID
- Family Structure during COVID
- Change in Family Structure
- Number of Minors in Care During COVID
- Adults in Care Over 18

Living environment:
- Type of Living Space During Confinement
- Property Ownership
- Square Meters of Living Space
- Light During Quarantine
- Ventilation During Quarantine
- Number of Rooms
- Number of Cohabitants
- Number of Cohabitants in the Same Room
- Outings During Quarantine
- Assessment of COVID Measures

Economic status:
- Employment Status Before Quarantine
- Employment Status During Quarantine
- Employment Status After Quarantine
- Change in Employment Status
- Working Hours Before Quarantine
- Occupation of the Person with the Highest Economic Contribution
- Net Monthly Income Before COVID
- Net Monthly Income During COVID
- Change in Net Monthly Income
- Financial Sufficiency and Situation During COVID
- Financial Sufficiency and Situation After COVID

Health impacts:
- Health Condition and Specific Needs
- Disability Degree
- Suicide Attempt

- COVID Diagnosis
- COVID Severity
- Isolation During COVID Diagnosis
- Family Member Diagnosed with COVID
- Family Member COVID Severity
- Co-living with a Diagnosed Family Member